\documentclass[prd,twocolumn,aps,psfig,nofootinbib,nobibnotes,superscriptaddress]{revtex4}
\usepackage{natbib}
\usepackage{graphicx}
\usepackage[figuresright]{rotating}
\usepackage{amssymb}
\usepackage{bm}
\usepackage{epsfig}
\usepackage{color}
\usepackage{amsmath}
\usepackage{appendix}
\usepackage{threeparttable}

\usepackage{multirow}
\usepackage{array}

\newcommand{\beq}{\begin{eqnarray}}
\newcommand{\eeq}{\end{eqnarray}}

\begin{document}

\title{The cosmological Mass Varying Neutrino model in the late universe}

\date{\today}

\author{Olga Avsajanishvili}
\affiliation{Evgeni Kharadze Georgian National Astrophysical Observatory, Tbilisi 0179, Georgia}


\begin{abstract}
The cosmological Mass Varying Neutrino (MaVaN) model is considered, where the interaction between a fermionic field and a scalar field with a Ratra–Peebles potential via a Yukawa coupling is investigated. 
Observational constraints on the flat and non-flat MaVaN models, as well as on the standard $\Lambda$CDM model, are derived from 32 $H(z)$ measurements using MCMC analysis.
Comparison with the $\Lambda$CDM model using the criteria $\Delta\chi^2_{\rm min}$, $AICc$, and $BIC$ shows no statistically significant improvement for MaVaN models. Deviations in the expansion history remain well below $1\sigma$, indicating that the $H(z)$ data alone do not provide sufficient constraining power to distinguish MaVaN models from the $\Lambda$CDM model. The non-flat MaVaN model reduces the tension between the $H_0$ value inferred from $H(z)$ data and Planck CMB measurement from $\sim 2\sigma$ in the $\Lambda$CDM framework to $\sim 1.1\sigma$. The discrepancy with the SH0ES measurement of $H_0$ is also reduced to below $1\sigma$, primarily due to the large uncertainties of the $H(z)$ data. 
\end{abstract}

\maketitle

\section{Introduction}\label{section:1}

The standard spatially flat $\Lambda$CDM cosmological
model~\citep{1984ApJ...284..439P,Peebles:2002gy,Caldwell2009,Silvestri:2009hh} (for a recent review, see~\citep{Lopez-Corredoira:2023jhh}) agrees well with various observational constraints obtained over the past decades~\citep{Planck:2018nkj,Planck:2018vyg,ACT:2020gnv, eBOSS:2020yzd}. 
 
Although the $\Lambda$CDM model remains the fiducial cosmological model, it exhibits several statistically significant tensions and anomalies with various datasets. The number of such discrepancies increases as more precise observational data become available (for recent comprehensive reviews, see~\citep{DiValentino:2021izs,Perivolaropoulos:2021jda,Moresco:2022phi,DiValentino:2022fjm,Abdalla:2022yfr, Peebles:2022akh,Hu:2023jqc,Vagnozzi:2023nrq}.
Among these discrepancies, one of the most prominent is the Hubble constant tension $H_0$, which manifests as a significant disagreement between Planck Cosmic Microwave Background (CMB) data~\cite{Planck:2018vyg} and local measurements from the Supernova $H_0$ for the Equation of State (SH0ES) collaboration~\citep{Riess:2021jrx}.

The presence of still unsolved problems in the $\Lambda$CDM model inspires and stimulates cosmologists to explore dark energy models that go beyond the $\Lambda$CDM model~\citep{DiValentino:2021izs,Abdalla:2022yfr,Khalife:2023qbu}.
The dynamical scalar field $\phi$CDM models are one of many alternatives to the $\Lambda$CDM model. In these models,  dark energy is represented in the form of a uniform cosmological scalar field that slowly varies in the present epoch~\citep{Ratra:1987rm,Ratra:1987aj, Wetterich:1987fk}. 
In these models, the energy density and pressure are time dependent functions under the assumption that the scalar field is described by the ideal barotropic fluid model.

One of the unresolved issues in modern cosmology is the coincidence problem~\citep{1992MPLA....7..563K,Comelli:2003cv,Velten:2014nra}, the essence of which is that at the present epoch, the density of dark energy is comparable to the energy density of dark matter, despite the fact that in the past they changed differently over time. This suggests that dark matter and dark energy could somehow interact with each other in the course of their evolution. Various interacting dark energy models have been proposed to provide a possible solution to the coincidence problem in the $\Lambda$CDM model~\citep{Fardon:2003eh,Farrar:2003uw,Cai:2004dk, Pavon:2005yx,Huey:2004qv,Berger:2006db,delCampo:2008sr, delCampo:2008jx}. 

In its turn, dark matter is comparable, in order of magnitude, to the energy density of neutrinos~\citep{Hannestad:2004nb,Dolgov:2008hz,Ivanchik:2024mqq}. Various experiments confirmed that neutrinos have mass~\citep{Tanabashi:2018oca}, but the origin of neutrino masses is still an unresolved issue. Stringent constraints on the sum of neutrino masses were obtained by different state-of-the-art cosmological measurements~\citep{Planck:2018nkj,Planck:2018vyg, Palanque-Delabrouille:2019iyz,eBOSS:2020yzd,ACT:2023kun, Tristram:2023haj,Naredo-Tuero:2024sgf,DESI:2024hhd, DESI:2024mwx,Jiang:2024viw,DESI:2025zgx,DESI:2025ejh}. 
The latest cosmological constraint on the upper limits on the sum of neutrino masses $\sum m_\nu(a_0)<0.0642~{\rm eV}$ was established through measurements of baryon acoustic oscillations (BAO) from Data Release 2 (DR2) by the Dark Energy Spectroscopic Instrument (DESI)~\citep{DESI:2025ejh} in galaxies and Lyman-$\alpha$ forest tracers. However, the Karlsruhe Tritium Neutrino Experiment (KATRIN), which directly measures the effective electron antineutrino mass, has reported an upper limit of $m_\nu(a_0)<0.45~{\rm eV}$ at the confidence level $90\%$~\cite{KATRIN:2024cdt}. This bound roughly corresponds to a constraint on the sum of neutrino masses
$\sum m_\nu < 1.35~{\rm eV}$ at the same confidence level.

Fardon, Nelson and Weiner elaborated the mechanism of Varying Mass Particles (VAMPs)~\citep{Fardon:2003eh}. They applied this mechanism in the context of neutrinos, as a result of which the Mass Varying Neutrino (MaVaN) model was created.
In this model, the neutrinos, {\it i.e.,} fermionic field, interact with the scalar field
via the Yukawa coupling. If the decoupled neutrino field is initially massless, then the coupling generates a mass of neutrinos, which subsequently changes over time~\citep{Farrar:2003uw}.
The MaVaN scenario is quite compelling, since it connects the origin of neutrino masses to dark energy, and solves the coincidence problem of the $\Lambda$CDM model. 
However, there are some problems that need to be addressed for the MaVaN scenario to be viable
~\citep{Afshordi:2005ym,PhysRevLett.96.061301, Kaplinghat:2006jk,Bjaelde:2007ki,Ichiki:2007ng, Bean:2007ny,Bean:2007nx}. One of these is related to the strong instability in the MaVaN model due to the negative square of the sound speed of the neutrinos-scalar field fluid~\citep{Afshordi:2005ym,Bean:2007ny,Bean:2007nx}. The coupling between neutrinos almost stops the evolution of the scalar field and triggers an accelerated expansion of the universe \citep{Wetterich:2007kr,Collodel:2012bp}. The exponential clustering of the neutrino occurs due to the exponential growth of scalar perturbations in the MaVaN model~\citep{Afshordi:2005ym,Kaplinghat:2006jk, Pettorino:2008ez,Mota:2008nj,Pettorino:2010bv, Nunes:2011mw,Ayaita:2012xm,Casas:2016duf}. The effect of the neutrinos-scalar field fluid in the MaVaN model on various cosmological observations has been studied in many papers, including~\citep{Brookfield:2005bz,Pettorino:2010bv,Brouzakis:2010md,Goh:2024exx}.

 Within the MaVaN model, \cite{Chitov:2009ph}  analyzed the thermal evolution of the universe and predicted its stable, metastable, and unstable phases. They found that the universe in the present epoch is below its critical temperature, and the state of the universe is similar to a supercooled liquid that has not yet crystallized: its high-temperature (meta)stable phase has become unstable, but the new low-temperature stable phase has not yet been reached.

\cite{Mandal:2019kkv} found that, depending on the choice of the scalar field potential, the acceleration of the universe occurs in an instability regime or in a stable regime in the MaVaN model. \cite{Mandal:2022yym} considered the interaction of the fermionic field and the scalar field with the exponential potential in the MaVaN model. They found that the current relict dark matter density is reached at fermion masses in the range of ${\rm1~GeV - 10^9~GeV}$ for the phase transition (after which the coupled neutrinos-scalar field fluid behaves as pressureless dark matter) temperature range of ${\rm 10~MeV - 10^7~GeV}$. 

In the framework of the MaVaN model, we explored the interaction between a fermionic field and a scalar field with the inverse power-law Ratra-Peebles potential during the late epoch of the universe. Our study follows the formalism elaborated in~\cite{Chitov:2009ph}; however, unlike~\cite{Chitov:2009ph}, which considered a single neutrino flavor ($N_F=1$), We perform numerical calculations for three flavors ($N_F=3$).

We constrain free parameters of the flat and non-flat MaVaN models, as well as those of the standard $\Lambda$CDM model. To compare these models, we employ the criteria $\Delta\chi^2_{\rm min}$, $AICc$, and $BIC$. Furthermore, using only $H(z)$ data, we investigate whether the MaVaN models can alleviate the $H_0$ tension relative to the $\Lambda$CDM scenario.

This paper is organized as follows: Section~\ref{section:2} presents the main equations for the MaVaN model, Section~\ref{section:3} considers the interaction of the fermionic and scalar fields with the Ratra-Peebles potential, Section~\ref{section:4} describes the observational constraints on the parameters of the flat and non-flat MaVaN models, Section~\ref{section:5} presents the results of the calculations, and Section~\ref{section:6} summarizes the conclusions. We use natural units, $c=k_B=1$.

\section{The general description of the MaVaN model} \label{section:2}
We assumed that the flat, homogeneous and isotropic universe is described by the Friedmann-Lemaître-Robertson-Walker spacetime metric $ds^2 = dt^2 -a^2(t)d{\bf x}^2$, here $t$ is a cosmic time, $a(t)$\footnote{Below, for simplicity of notation, we omit the explicit dependence of the scale factor on time, while implying the dependence of the scale factor on time.} is the scale factor (normalized to unity in the present epoch, $a_0 \equiv a(t_0)=1$).

The Euclidean action of the scalar field is defined as
\begin{equation}\label{eq:Action_Scalar}
S_B^E=\int_0^\beta d\tau\int a^3d^3x\left[\frac{1}{2}\left(\frac{\partial\phi}{\partial\tau}\right)^2+\frac{(\nabla\phi)^2}{2a^2}+V(\phi)\right],
\end{equation}
here $\tau=it$ is the Euclidean time, $\int d^3x = V_{com}$ is the comoving volume; $a^3\int d^3x = V_{phys}$ is the physical volume;  $V(\phi)$ is the potential of the scalar field.

The Euclidean action for the Dirac field is given by
\begin{equation}\label{eq:Action_Dirac}
S_D^E=\int_0^\beta d\tau\int a^3\,d^3x\,\bar{\psi}({\bf x},\tau)\hat{D}(\phi)\psi({\bf x},\tau),
\end{equation}
where the Dirac operator is 
\begin{equation}\label{eq:Dirac_operator}
\hat{D}(\phi) = \gamma^\circ\frac{\partial}{\partial \tau}-\frac{\iota{\gamma}}{a}\cdot\nabla+g\phi({\bf x},\tau)-\mu \gamma^\circ,
\end{equation}
here $g$ is the Yukawa coupling; $\mu$ is the chemical potential; following~\citep{Chitov:2009ph}, we set $g=1$ and $\mu=0$ in our analysis.

The fermionic contribution to the pressure is defined as
\begin{equation}\label{eq:Dirac pressure} 
P_D=-P_0+\frac{1}{3\pi^2}\int_0^\infty\frac{k^4 dk}{\epsilon(k)}\bigg[n_F(\epsilon_-)+n_F(\epsilon_+)\bigg],
\end{equation}
here $P_0=2\int d^3 k\epsilon(k)/(2\pi)^3$ is the 
contribution of the vacuum to the pressure; $\epsilon(k)=\sqrt{k^2+m_\nu^2}$, $\epsilon_{\pm}=\epsilon\pm\mu$,  $k$ is the Fermi momentum, $m_\nu$ is the fermionic mass; $n_F(x) \equiv (e^{\beta x}+1)^{-1}$ is the Fermi-Dirac distribution function; $\beta=1/T_{\nu}$ is the inverse temperature, where the fermion temperature evolves as $T_\nu=T_{\nu0}/a$  and  $T_{\nu0} = 1.9454~K=2.35\cdot10^{-4}~{\rm eV}$ is the temperature of the fermions in the present epoch.

The total Euclidean action describing the interacting scalar and fermion fields is
\begin{equation}\label{eq:Action_Scalar_Dirac}
S_{BD}^E = S_B^E + S_D^E =\Big |_{m_\nu=0} + g\int_0^\beta d\tau\int a^3
 d^3x \phi\bar{\psi}\psi,
\end{equation}
where the Yukawa interaction term $g\,\phi\,\bar{\psi}\psi$ gives rise to a dynamically generated fermion mass. This mass originates from the spontaneous breaking of the chiral symmetry in the Dirac sector of the Lagrangian density corresponding to Eq.~(\ref{eq:Action_Scalar_Dirac}) and evolves dynamically in response to the scalar field.

The path integral for the partition function in the interaction of
the scalar field with the fermionic field is defined as
\begin{equation}\label{eq:Path_integral}
\mathcal{Z} = \int\mathcal{D}\phi\,\mathcal{D}\bar{\psi}\,\mathcal{D}\psi\,e^{-S_{BD}^E}.
\end{equation}

The Grassmann fields can be formally integrated~\citep{Kapusta:2006pm} as 
\begin{equation}\label{eq:Grassmann_field}
\mathcal{Z} = \int\mathcal{D}\phi e^{-S_{BD}^E} = \int\mathcal{D}\phi \exp(-S_{BD}^E+\log {\rm Det}\hat{D}(\phi)).
\end{equation}

The total thermodynamic potential of the coupled fermionic and scalar fields can be written as
\begin{equation}\label{eq:Total_thermodynamic_potential} 
V_{\phi\nu} = V(\phi) + V_\nu(\phi) = V(\phi) - \frac{2N_F}{3\pi^2\beta^4}\int^{\infty}_{\overline{\phi}} \frac{(x^2-\overline{\phi}^2)^{3/2}}{e^x+1}dx,
\end{equation}                    
here $N_F$ is a number of neutrino flavors (we take $N_F=3$); $\overline{\phi} = g\beta\phi$~\citep{Chitov:2009ph} is a dimensionless scalar variable.
This potential  can be found in the saddle-point approximation, minimizing the path integral, Eq.~(\ref{eq:Grassmann_field}). The saddle-point approximation coincides with the minimum condition of the total thermodynamic potential at equilibrium for the fixed temperature and the chemical potential values at the critical point $a_{cr}$  
\begin{equation}\label{eq:Condition of minimum} 
\frac{\partial V_{\phi\nu} }{\partial \phi}\Bigg|_{\nu, \beta,\phi = \phi_{cr}} = 0,~~~~~\frac{\partial^2 V_{\phi\nu}}{\partial \phi^2}\Bigg|_{\nu, \beta,\phi = \phi_{cr}} > 0.
\end{equation}
As a consequence, neutrinos acquire mass at this critical point, and the corresponding condition can be written as
\begin{equation}\label{eq:Mass at the critical point} 
\frac{\partial V_{\phi\nu}}{\partial \phi}\Bigg|_{\phi = \phi_{ cr}} = \frac{\partial V(\phi)}{\partial \phi}\Bigg|_{\phi = \phi_{ cr}} + \frac{\partial V_\nu(\phi)}{\partial \phi}\Bigg|_{\phi = \phi_{ cr}},
\end{equation}
here $\phi(a_{ cr})=\phi_{cr}$ is the average of the bosonic field, $\phi_{cr}=\langle \phi\rangle$, and the fermionic mass is defined as $m_\nu = g\phi_{cr}$\footnote{Although microscopic theory is formulated in terms of the mass of a single-particle, the quantity constrained by cosmological observations is the total neutrino mass $\sum m_\nu = N_Fg\langle\phi\rangle =  N_F g \phi_{\rm cr} = N_F m_\nu$. This distinction ensures a consistent theoretical formulation and allows for a direct comparison with observational bounds on neutrino masses.}~(\cite{Chitov:2009ph}).

Eq.~(\ref{eq:Mass at the critical point}) can be represented as
\begin{equation}\label{eq:Potential at the critical point} 
V'(\phi)\big|_{\phi=\phi_{cr}}+g\rho_s=0,
\end{equation}
where $\rho_s$ is the scalar fermionic density (or the chiral density) that is given by
\begin{equation}\label{eq:chiral density} 
\rho_s=\frac{m_\nu}{\pi^2}\int_0^\infty\frac{dkk^2}{\epsilon(k)}\bigg[n_F(\epsilon_+)+n_F(\epsilon_-)-1)\bigg].
\end{equation}

\section{Interaction of the fermionic field and the scalar field with the  Ratra-Peebles potential}\label{section:3}

\subsection{Scalar field with the Ratra-Peebles potential}

\begin{figure}[!ht]
\begin{center}
\includegraphics[width=\columnwidth]{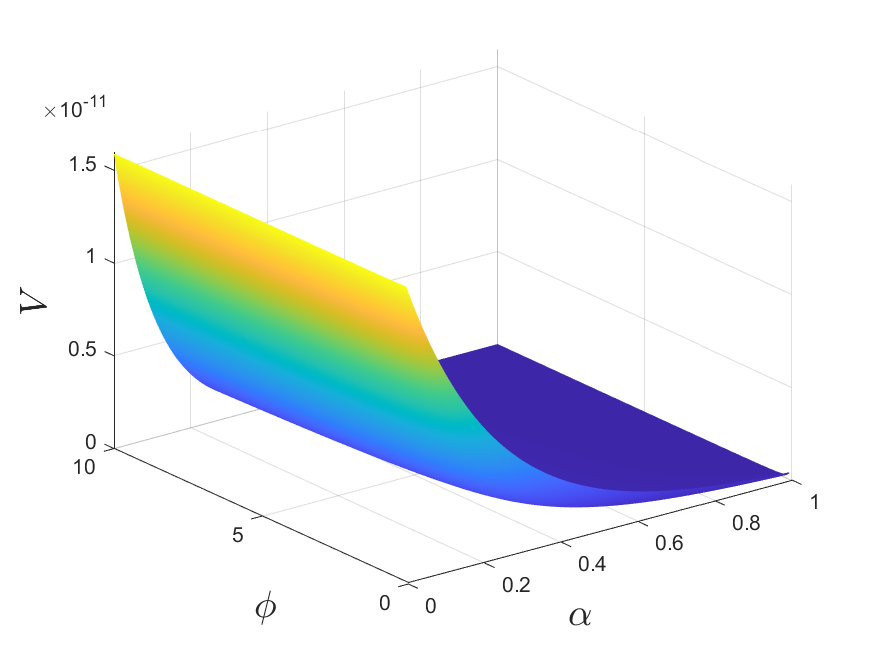}
\end{center}
 \caption {A three-dimensional representation of the Ratra-Peebles potential for $M=2\cdot 10^{-3}$ eV.}
 \label{fig:f1}
 \end{figure}

We studied the interaction of the fermionic field and the scalar field with the inverse power law Ratra-Peebles potential~\citep{Ratra:1987rm, Ratra:1987aj}, which has a form
\begin{equation}\label{eq:Ratra-Peebles potential}
V(\phi)=\frac{M^{\alpha+4}}{\phi^\alpha},
\end{equation}
here $M$ and $\alpha$ are positive model parameters. The parameter $M$ is the scale mass of the Ratra-Peebles potential; the parameter $\alpha$ defines the steepness of the potential, for $\alpha=0$, the scalar field $\phi$CDM model reduces to the standard $\Lambda$CDM model. A three-dimensional representation of the Ratra-Peebles potential is shown in Fig.~\ref{fig:f1}

The total thermodynamic potential of the coupled fermionic and scalar fields with the Ratra-Peebles potential at the critical point $a_{cr}$ reads as follows 
\begin{equation}\label{eq:General_thermodynamic_potential} 
V_{\phi\nu} = \frac{M^{\alpha+4}}{\phi^\alpha} - \frac{2N_F}{3\pi^2\beta^4}\int^{\infty}_{\overline{\phi}} \frac{(x^2-\overline{\phi}^2)^{3/2}}{e^x+1}dx.
\end{equation}

\subsection{Basic equations describing the neutrinos-scalar field fluid}
Before interaction with the scalar field, neutrinos are in the free-streaming regime, being massless and relativistic. After the critical point, neutrinos become non-relativistic, {\it i.e.,} the  chiral density of neutrinos scales as $\rho_\nu \propto a^{-3}$ for $a \geq a_{cr}$. During this period of time, the formation of the neutrinos-scalar field fluid occurs.

The neutrinos-scalar field fluid is characterized by the total effective potential consisting of the Ratra-Peebles potential and the additional component connected to neutrinos
\begin{equation}\label{eq:Effective potential}
V_{ couple} = \frac{M^{\alpha+4}}{\phi^\alpha} +\phi \rho_{ cr}\Bigg(\frac{a_{cr}}{a}\Bigg)^3,
\end{equation}
here $\rho_{ cr}=\alpha(\Delta_{cr}/\nu)^{(\alpha+1)}M^3$ is a value of the chiral density at the critical point;
$\Delta_{ cr}=\bigg(\frac{\sqrt{2}\nu^{\nu}e^{-\nu}}{\alpha \pi^{3/2}}\bigg)^{1/(\alpha+4)}$, with $\nu=\alpha+5/2$, $M=(\nu^{\alpha}\rho_{\phi\nu0})^{(\alpha+1)/(\alpha+4)}\Delta_{ cr}^{-\alpha}T_{\nu0}^{-3\alpha/(\alpha+4)}$~\citep{Chitov:2009ph}; $\rho_{\phi\nu0}\approx2.76\cdot 10^{-11}h^2~{\rm eV^4}$ is the density of the neutrinos-scalar field fluid in the present epoch, with $h=0.674$~\citep{Planck:2018vyg}.

The first Friedmann equation and the scalar field Klein-Gordon equation of motion of the neutrinos-scalar field fluid are respectively read as
\begin{multline} \label{eq:First_Friedmann}
E = \Bigg(\Omega_{m0} a^{-3} + \Omega_{k0} a^{-2} + \frac{1}{\rho_{ cr0}}\Bigg(\frac{\dot{\phi}^2}{2} + \frac{M^{\alpha+4}}{\phi^\alpha} + \\\phi \rho_{ cr}\Bigg(\frac{a_{cr}}{a}\Bigg)^3 \Bigg)\Bigg)^{1/2},
\end{multline}

\begin{multline}\label{eq:Klein-Gordon}
\ddot{\phi}+3H{\dot\phi} - \frac{\alpha M^{\alpha+4}}{\phi^{\alpha+1}} + \rho_{cr}\Bigg(\frac{a_{cr}}{a}\Bigg)^3=0,
\end{multline}
here $E = H/H_0$ is the normalized Hubble parameter, $H$ is the Hubble parameter, $H_0$  the value of the Hubble parameter in the present epoch,  $H_0 = 100\cdot h~{\rm km~s^{-1}Mpc^{-1}}$; $\Omega_{m0}$ and $\Omega_{k0}$ are the  parameters of the density matter and curvature in the present epoch,  $\Omega_{m0} = 0.315$~\cite{Planck:2018vyg}; $\rho_{ cr0}\approx4.31\cdot 10^{-11}h^2~{\rm eV^4}$ is the critical energy density in the present epoch.

The mass of the scalar field with the Ratra-Peebles potential is defined as
\begin{equation}\label{eq:Mass of the scalar field} 
m_\phi = \Bigg(\frac{\partial^2 V}{\partial \phi^2}\Bigg)^{1/2}\bigg|_{\phi = \phi_{ cr}}=\Big(\alpha(\alpha+1)M^{\alpha+4}\phi^{-(\alpha+2)}\Big)^{1/2}.
\end{equation} 

\section{Observational constraints on free parameters of flat and non-flat MaVaN and $\Lambda$CDM models}\label{section:4}
We constrained the free parameters
${\bf p}=(H_0, \alpha, \Omega_{m0}, \sum m_\nu)$,
${\bf p}=(H_0, \alpha, \Omega_{m0}, \sum m_\nu, \Omega_{k0})$,
and ${\bf p}=(H_0, \Omega_{m0})$
of the flat MaVaN, non-flat MaVaN, and $\Lambda$CDM models, respectively. For this analysis, we used 32 $H(z)$ measurements in the redshift range
$0.07 \leq z \leq 1.965$, as compiled in~\citep{Cao:2023eja}. Among these, 15 measurements are correlated \citep{2012JCAP...08..006M, Moresco:2015cya, Moresco:2016mzx}. The covariance matrix for the correlated data is publicly available at https://gitlab.com/mmoresco/CCcovariance/.

In our analysis, we adopted the following priors for the free parameters (see Table~\ref{table:priors}), where the sum of the neutrino masses $\sum m_\nu$ is expressed in units of $\rm eV$ and the Hubble parameter in the present epoch $H_0$ in units of $\rm km~s^{-1}Mpc^{-1}$.

\begin{table} [!hbp]
\caption{Priors for free parameters of the models.}
 \label{table:priors}
 \centering
\begin{tabular}{cc}
\hline\noalign{\smallskip}
$\rm Parameter$   &     $\rm Priors$ \\
\hline
$H_0$ & $[\rm None,~ None]$ \\
$\alpha$ &  $[0.0001,0.04]$ \\
$\Omega_{m0}$ &  $[0.15,0.6]$  \\
$\sum m_\nu$ &  $[0.0001,0.6]$ \\
$\Omega_{k0}$ &  $[-0.6,0.6]$ \\
\noalign{\smallskip}\hline
\end{tabular}
  \end{table}
 
The posterior distributions of the parameters for the flat and non-flat MaVaN models and the $\Lambda$CDM model were sampled using the Markov Chain Monte Carlo (MCMC) method via the Cobaya framework.
 \citep{Torrado:2020dgo}. In addition, we applied the PYTHON GETDIST package \citep{Lewis:2019xzd} to create plots and analyze the MCMC results. Chain convergence was rigorously validated using the Gelman-Rubin diagnostic, achieving $R-1=0.0226$,  $R-1=0.0362$ and $R-1=0.0151$, for flat  MaVaN, non-flat MaVaN and $\Lambda$CDM models, respectively, which satisfies the conventional threshold of $R-1 \leq 0.1$ \citep{Gelman:1992} and indicates robust sampling of the target distribution.

To statistically compare the flat and non-flat MaVaN models with the $\Lambda$CDM model, which differ in complexity, and to identify the model that best balances goodness-of-fit and parameter parsimony, we employed the corrected Akaike Information Criterion ($AICc$)~\cite{10.1093/biomet/76.2.297}
\begin{equation}
AICc = \chi^2_{\min} + 2k + \frac{2k(k+1)}{n - k - 1},
\end{equation}
where $\chi^2_{\min}$ is the minimum chi-squared value, $k$ denotes the number of free parameters, $n$ is a number of data points.
Due to the limited sample size ($n=32$, $n/k = 6.4-16 < 40$), the corrected Akaike Information Criterion ($AICc$) was used to properly correct for the known bias of the standard $AIC$ toward over-parameterized models in small-sample cases.
~\cite{10.1093/biomet/76.2.297}. To verify the robustness of the result, the Bayesian Information Criterion ($BIC$)~\cite{Schwarz1978} was also calculated
\begin{equation}
BIC = \chi^2_{\min} + k\ln {n}.
\end{equation}
Both criteria are evaluated based on the differences $\Delta AICc$ and $\Delta BIC$ relative to the best model, as well as the model probability weights (Akaike weights) $w_i$~\cite{burnham2002} 
\begin{equation}
w_i=\frac{\exp{(-\Delta_i/2)}}{\sum_{j=1}^R(-\Delta_j/2)},
\end{equation}
here $\Delta_i=AICc_i-{\rm min}(AICc)$ is the difference in the value of $AICc$ for model $i$ relative to the best model (with the minimum $AICc$); $R$ is the number of models compared; the sum in the denominator is taken over all $R$ models. These weights are interpreted as the probability that model $i$ is the best (in terms of minimizing information loss) among the set of candidate models, given that one of them is true.

The parameters $\Delta\chi^2_{\min}=\chi^2_{\min,i}-\rm min(\chi^2_{\min})$ and the minimum reduced chi-square, $\chi^2_{\min}/{\rm dof}$ are used to assess the goodness of fit, where ${\rm dof}=n-k$ denotes the number of degrees of freedom.

\section{Results and discussions}\label{section:5}
\subsection{Features of the flat MaVaN model}

 To study the evolution of the neutrinos-scalar field fluid, we jointly numerically integrated the first Friedmann equation,  Eq.~(\ref{eq:First_Friedmann}), and the scalar field equation of motion, Eq.~(\ref{eq:Klein-Gordon}). We perform numerical calculations for the flat MaVaN model, taking into account $\Omega_{k0}=0$ in Eq.~(\ref{eq:First_Friedmann}). 
 
\begin{figure}[!ht]
\begin{center}
\includegraphics[width=\columnwidth]{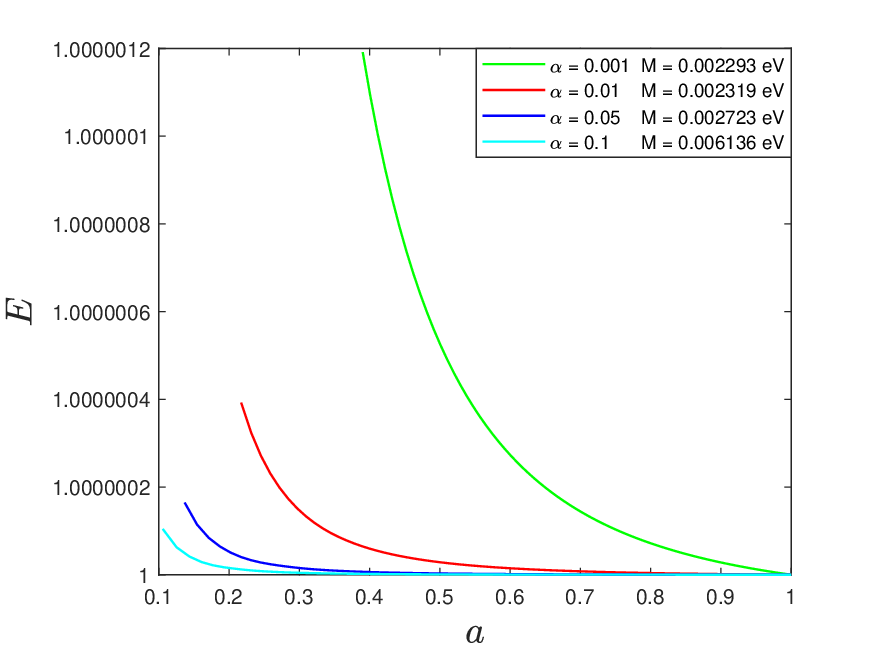}
\end{center}
\caption{The normalized expansion rate of the universe in the flat MaVaN model, taking into account $N_f=3$, for different values of the model parameter $\alpha$.}
 \label{fig:f2}
 \end{figure}

We investigate the influence of the neutrinos - scalar field fluid on the expansion rate of the universe. The universe expands more slowly with an increasing value of the parameter $\alpha$ and vice versa for all values of the scale factor, but only in the present epoch the expansion of the universe occurs equally regardless of the value of the model parameter $\alpha$ (see Fig.~\ref{fig:f2}). In contrast to this fact, in the case of the scalar field with the Ratra-Peebles potential that does not interact with neutrinos, with an increase in the value of the parameter $\alpha$, the scalar field contribution becomes stronger and, as a consequence, the expansion of the universe occurs faster~\citep{Avsajanishvili:2014jfy}. In the case of interaction between the scalar field and neutrinos and the formation of the neutrinos-scalar field fluid, with an increase in the value of the parameter $\alpha$, the sum of the neutrino masses (see Table II,~\citep{Chitov:2009ph}) increases and, accordingly, the scalar field weakens, leading to a slowdown in the expansion of the universe.
 
\begin{figure}[!ht]
\begin{center}
\includegraphics[width=\columnwidth]{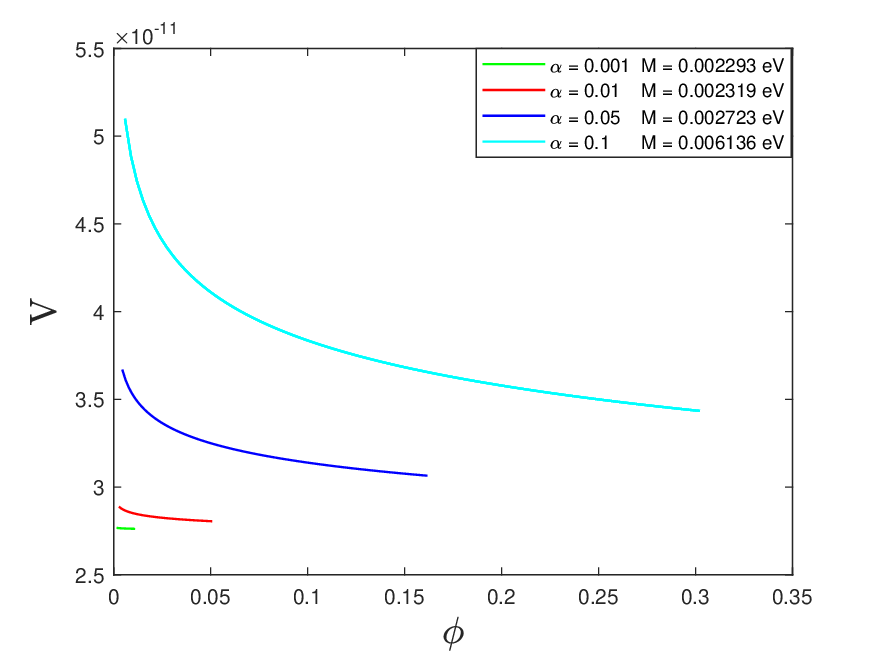}
\end{center}
 \caption{The mutual influence of the sum of neutrino masses and the scalar field Ratra-Peebles potential, for different values of the model parameter $\alpha$.}
 \label{fig:f3}
 \end{figure}

 By studying the mutual influence of the value of the sum of neutrino masses and the value of the scalar field potential, we found that with an increase in the value of the model parameter $\alpha$, the value of the Ratra-Peebles scalar field potential also increases (see Fig.~\ref{fig:f3}). This happens because the scalar field potential is defined in a certain range of neutrino masses, which depends on the value of the model parameter $\alpha$.
 
\begin{figure}[ht!]
\begin{center}
\includegraphics[width=\columnwidth]{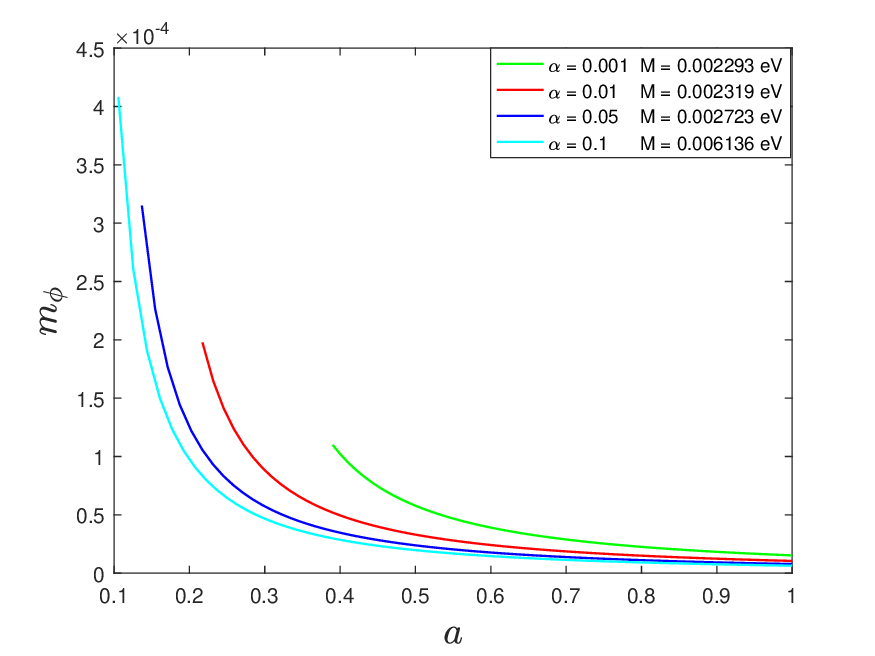}
\end{center}
 \caption {The dependence of the mass of the scalar field with the Ratra-Peebles potential on the model parameter $\alpha$.}
 \label{fig:f4}
 \end{figure}

We explored the evolution of the mass of the scalar field with the Ratra-Peebles potential from the critical point $a_{cr}$ to the present epoch $a_0$ depending on the model parameter $\alpha$.
With an increasing value of the model parameter $\alpha$, the mass of the scalar field decreases, while all these tracks converge to a zero mass value in the present epoch (see Fig.~\ref{fig:f4}).

\subsection{Observational constraints on free model parameters}

Using 32 $H(z)$ measurements, we constrained the free parameters of the flat and non-flat MaVaN models, as well as of the standard $\Lambda$CDM model, by means of a MCMC analysis. The resulting best-fit values and the corresponding $1\sigma$ and $2\sigma$ confidence level intervals are summarized in Tables~\ref{table:2} and \ref{table:3}.
For the flat MaVaN model, the best-fit Hubble constant is relatively low, $H_0 = 63.34^{+3.61}_{-2.18}$ km s$^{-1}$ Mpc$^{-1}$  at $1\sigma$ confidence level ($H_0 = 63.34^{+7.22}_{-4.36}$ km s$^{-1}$ Mpc$^{-1}$  at $2\sigma$ confidence level), while the matter density parameter in
the present epoch, $\Omega_{m0}=0.343^{+0.050}_{-0.053}$ at $1\sigma$ confidence level, is slightly higher than in the $\Lambda$CDM model.
The sum of the neutrino masses $\sum m_\nu$ and parameter $\alpha$ are small and consistent with zero within both $1\sigma$ and $2\sigma$ confidence levels.
In the non-flat MaVaN model, the best-fit Hubble constant increases to $H_0 = 67.52^{+6.05}_{-5.01}$ km s$^{-1}$ Mpc$^{-1}$  at $1\sigma$ confidence level ($H_0 = 67.52^{+12.10}_{-10.02}$ km s$^{-1}$ Mpc$^{-1}$  at $2\sigma$ confidence level), while the  matter density parameter in the present epoch $\Omega_{m0} = 0.386^{+0.070}_{-0.095}$ at $1\sigma$ confidence level exhibits a higher central value than both the flat MaVaN and $\Lambda$CDM models. The curvature parameter in the present epoch $\Omega_{k0} = -0.199^{+0.196}_{-0.186}$ at $1\sigma$ confidence level ($2\sigma$ range includes nearly flat values) suggests a mildly closed universe, however, the large uncertainty implies that spatial flatness cannot be excluded.  The parameters $\sum m_\nu$ and $\alpha$ remain consistent with those obtained in the flat MaVaN case within $1\sigma$, indicating that the $H(z)$ data provide only weak constraints on these quantities.
For the $\Lambda$CDM model, the best-fit Hubble constant is higher, $H_0 = 68.59^{+2.40}_{-2.96}$ km s$^{-1}$ Mpc$^{-1}$  at $1\sigma$ confidence level ($H_0 = 68.59^{+4.80}_{-5.92}$ km s$^{-1}$ Mpc$^{-1}$  at $2\sigma$ confidence level), while the matter density parameter in the present epoch is slightly lower, $\Omega_{m0} = 0.315^{+0.050}_{-0.040}$ at $1\sigma$ confidence level, compared to the MaVaN models.

We investigated whether the MaVaN cosmological models constrained by the $H(z)$ data can alleviate the tension $H_0$ between early- and late-universe measurements. Within the standard $\Lambda$CDM framework, the Planck CMB and SH0ES measurements correspond to $H_0^{\rm Planck} = 67.4 \pm 0.5$ km s$^{-1}$ Mpc$^{-1}$~\cite{Planck:2018vyg} and $H_0^{\rm SH0ES} = 73.04 \pm 1.04$ km s$^{-1}$ Mpc$^{-1}$~\cite{Riess:2021jrx}, both quoted at the $1\sigma$ confidence level, which corresponds to a tension of approximately $5.4\sigma$. 
For the non-flat MaVaN model, the inferred Hubble constant $H_0$ at the $1\sigma$ confidence level is statistically consistent with the Planck CMB result\footnote{
We note that the quoted Planck CMB constraint on $H_0$ is derived assuming the standard $\Lambda$CDM cosmological model. A full CMB analysis within the MaVaN framework would generally lead to different inferred cosmological parameters and could modify the level of statistical agreement with the Planck CMB result. Such an analysis is beyond the scope of the present work, and here the Planck CMB  result is used only as a reference benchmark for comparison.}, with the discrepancy reduced to a negligible level due to the relatively large uncertainties inferred from the $H(z)$ data. The difference from the SH0ES measurement is also weakened and remains within the $1\sigma$ permits a broader range of $H_0$, leading to improved nominal agreement with early-universe constraints. Nevertheless, this apparent mitigation of the $H_0$ tension should be interpreted with caution, as it primarily arises from enlarged confidence intervals rather than from a statistically significant shift of the best-fit value toward the local SH0ES measurement. Consequently, the tension is reduced but not fully resolved.
In contrast, the flat MaVaN model favors a lower value of $H_0$ at the $1\sigma$ confidence level, which is in mild tension with Planck CMB result at the $\sim 1.4\sigma$ confidence level and significantly lower than the SH0ES measurement, corresponding to a discrepancy of approximately $3\sigma$. This indicates that the flat MaVaN scenario is disfavored by the $H(z)$ data in the context of current constraints on $H_0$.
For the standard $\Lambda$CDM model, the inferred value of $H_0$ at the $1\sigma$ confidence level lies between those obtained for the flat and non-flat MaVaN models. It remains consistent with the Planck CMB result and partially alleviates the tension with the SH0ES measurement relative to the flat MaVaN case. However, owing to its comparatively smaller uncertainties, a residual discrepancy with SH0ES measurement persists at the $\sim 1.1\sigma$ confidence level, indicating that the $\Lambda$CDM model does not fully resolve the $H_0$ tension.

We presented the best-fit Hubble parameter values $H(z)$ together with the $1\sigma$, $2\sigma$, and $3\sigma$ confidence regions and the $1\sigma$ error bars for the independent and correlated $H(z)$ data in the flat and non-flat MaVaN models, as well as in the standard $\Lambda$CDM model. These results are shown in Fig.~\ref{fig:f5} (left and right panels), respectively.
 The $H(z)$ data have large uncertainties, and almost all of the data points lie within the $3\sigma$ confidence region of the best-fit $H(z)$  values for both the flat and non-flat MaVaN models, as well as for the standard $\Lambda$CDM model. 
A comparison of the best-fit $H(z)$ curves reveals a mild redshift-dependent behavior between the MaVaN and $\Lambda$CDM models. In the flat MaVaN scenario, the expansion rate exceeds that of 
$\Lambda$CDM model at low redshifts ($z\leq0.5$), becomes comparable at intermediate redshift values, and falls below it at higher redshifts. In contrast, the non-flat MaVaN model exhibits a different redshift-dependent pattern, with $H(z)$ exceeding the prediction of the $\Lambda$CDM model  at low redshifts ($z\leq0.35$), lying below it at intermediate redshifts, and becoming consistent again at higher redshifts. From a physical perspective, these behaviors indicate that MaVaN models modify the redshift dependence of the cosmic expansion rate, leading to small redshift-dependent deviations from the $\Lambda$CDM model, particularly at low redshifts. The sign change of the deviations as a function of $z$ may reflect the dynamical nature of the neutrino–scalar field coupling that drives cosmic acceleration in the MaVaN framework. However, all such deviations are significantly smaller than the 1$\sigma$ confidence level, and therefore do not represent statistically significant departures from the standard $\Lambda$CDM model.

We obtained one-dimensional likelihoods and two-dimensional $1\sigma$, $2\sigma$, and $3\sigma$ confidence-level contours for the parameters of the flat and non-flat MaVaN models using $H(z)$ measurements. The results for the flat and non-flat cases are shown in the left and right panels of Fig.~\ref{fig:f6}, respectively.

\begin{table*}[!ht]
\centering
\begin{threeparttable}
\caption{One-dimensional posterior best-fit parameters with $1\sigma$ uncertainties for flat and non-flat MaVaN models and the $\Lambda$CDM model from $H(z)$ data.}
\label{table:2}
\begin{tabular}{lcccccc}
\hline\noalign{\smallskip}
Model & {$H_0$\tnote{(a)}} & $\Omega_{m0}$ &  $\Omega_{k0}$ & {$\sum m_{\nu}$\tnote{(b)}} & $\alpha$  \\ 
\noalign{\smallskip}
\hline
\noalign{\smallskip}
Flat MaVaN & $63.34^{+3.61}_{-2.18}$ &  $0.343^{+0.050}_{-0.053}$ & $ - $ & $0.27^{+0.21}_{-0.17}$ & $0.007^{+0.005}_{-0.004}$ \\
\noalign{\smallskip}
\hline
\noalign{\smallskip}
Non-flat MaVaN & $67.52^{+6.05}_{-5.01}$ & $0.386^{+0.070}_{-0.095}$ & $-0.199^{+0.196}_{-0.186}$ & $0.27^{+0.21}_{-0.18}$ & $0.009^{+0.009}_{-0.006}$  \\
\noalign{\smallskip}
\hline
\noalign{\smallskip}
$\Lambda$CDM & $68.59^{+2.40}_{-2.96}$ & $0.315^{+0.050}_{-0.040}$ & $-$ & $-$ & $-$ \\ 
\noalign{\smallskip}\hline
\end{tabular}
\begin{tablenotes}
\item[(a)] km~s$^{-1}$Mpc$^{-1}$
\item[(b)] eV
\end{tablenotes}
\end{threeparttable}
\end{table*}

\begin{table*}[!ht]
\centering
\begin{threeparttable}
\caption{One-dimensional posterior best-fit parameters with $2\sigma$ uncertainties for flat and non-flat MaVaN models and the $\Lambda$CDM model from $H(z)$ data.}
\label{table:3}
\begin{tabular}{lcccccc}
\hline\noalign{\smallskip}
Model & {$H_0$\tnote{(a)}} & $\Omega_{m0}$ &  $\Omega_{k0}$ & {$\sum m_{\nu}$\tnote{(b)}} & $\alpha$  \\ 
\noalign{\smallskip}
\hline 
\noalign{\smallskip}
Flat MaVaN & $63.34^{+8.34}_{-3.17}$ &  $0.343^{+0.090}_{-0.108}$ & $ - $ & $0.27^{+0.31}_{-0.25}$ & $0.007^{+0.013}_{-0.005}$ \\
\noalign{\smallskip}
\hline 
\noalign{\smallskip}
Non-flat MaVaN & $67.52^{+10.76}_{-7.01}$ & $0.386^{+0.104}_{-0.170}$ & $-0.199^{+0.404}_{-0.334}$ & $0.27^{+0.31}_{-0.26}$ & $0.009^{+0.018}_{-0.008}$ \\
\noalign{\smallskip}
\hline
\noalign{\smallskip}
$\Lambda$CDM & $68.59^{+5.26}_{-4.68}$ & $0.315^{+0.078}_{-0.085}$ & $-$ & $-$ & $-$ \\
\noalign{\smallskip}\hline
\end{tabular}
\begin{tablenotes}
\item[(a)] km~s$^{-1}$Mpc$^{-1}$
\item[(b)] eV
\end{tablenotes}
\end{threeparttable}
\end{table*}

Table~\ref{table:4} summarizes the goodness-of-fit statistics and model comparison criteria for the flat and non-flat MaVaN models compared to the standard $\Lambda$CDM model, based on the $H(z)$ data. The minimum $\chi^2$ values for all models are very close, with $\chi^2_{\min} = 14.77$ for $\Lambda$CDM and slightly higher values for the flat and non-flat MaVaN models. The differences $\Delta\chi^2_{\min} = \chi^2_{\min}(\rm MaVaN) - \chi^2_{\min}(\Lambda \rm CDM)$, are small ($|\Delta\chi^2_{\min}|<1$) and do not indicate a statistically significant improvement. The reduced chi-square values, $\chi^2_{\min}/\mathrm{dof} < 1$, reflect the large uncertainties in the $H(z)$ data and the limited discriminating power of these measurements.

The comparison based on the information criteria $AICc$ and $BIC$ clearly favors the standard $\Lambda$CDM model, which achieves the lowest values $AICc = 19.18$ and $BIC = 21.70$, and the highest probability weight $w = 0.93$. In contrast, the flat and non-flat MaVaN models receive negligible weights, $w = 0.06$ and $0.02$, indicating that the additional parameters in these models do not provide a meaningful improvement in the fit. Although both MaVaN models produce acceptable fits, their higher penalties $AICc$ and $BIC$  due to additional parameters reduce their relative statistical support.

Specifically, both MaVaN models exhibit positive values of 
$\Delta AICc = AICc_{\rm MaVaN} - AICc_{\Lambda \rm CDM}$ and 
$\Delta BIC = BIC_{\rm MaVaN} - BIC_{\Lambda \rm CDM}$, with 
$\Delta AICc > 5$ and $\Delta BIC > 7$ (see Table~\ref{table:4}). 
Following~\cite{burnham2002}, these differences indicate essentially no support for the MaVaN models compared to the $\Lambda$CDM model, as they result from the inclusion of additional free parameters rather than a significant reduction in $\chi^2_{\min}$. The more stringent $BIC$, which imposes a stronger penalty for model complexity, particularly disfavors the non-flat MaVaN model. 
This demonstrates that the added complexity of the MaVaN scenarios does not lead to a statistically significant improvement in fitting the data and is therefore not justified.

\begin{table*}[!ht]
\centering
\begin{threeparttable}
\caption{Goodness-of-fit and model comparison criteria for flat and non-flat MaVaN models and the $\Lambda$CDM model from $H(z)$ data.}
\label{table:4}
\begin{tabular}{lccccccccc}
\hline\noalign{\smallskip}
$\rm Model$ & $ \rm \chi^2_{min}$ & $\rm \Delta\chi^2_{min}$ & $\rm dof$ & $\rm \chi^2_{min}/dof$ & $\rm AICc$ & $\rm BIC$ & $\Delta\rm AICc$ & $\Delta\rm BIC$ & $w$\\
\hline 
Flat MaVaN & $15.25$ & $0.48$ & $28$ & $0.54$ & $24.73$ & $29.11$ & $5.55$ & $7.41$ & $0.06$  \\
\hline
Non-flat MaVaN & $15.18$ & $0.41$ & $27$ & $0.56$ & $27.49$ & $32.51$ & $8.31$ & $10.81$ & $0.02$\\
\hline   
$\Lambda$CDM & $14.77$ & $0.00$ & $30$ & $0.49$ & $19.18$ & $21.70$ & $0.00$ & $0.00$ & $0.93$ \\
\hline
\end{tabular}
\end{threeparttable}
\end{table*}

\section{Conclusion}\label{section:6}

Studying the interaction of the fermionic field (assuming that a number of neutrino flavors $N_F=3$) and the scalar field with the Ratra-Peebles potential, we found that with an increase in the value of the model parameter $\alpha$: (i) the expansion of the universe occurs more slowly  for all values of the scale factor, but only in the present epoch the expansion of the universe occurs equally regardless of the value of the model parameter $\alpha$;  (ii) the mass of the scalar field decreases, while all these tracks converge to a zero mass value in the present epoch.

\begin{figure*}[!ht]
\centering
\begin{minipage}{.49\textwidth}
\includegraphics[width=\linewidth]{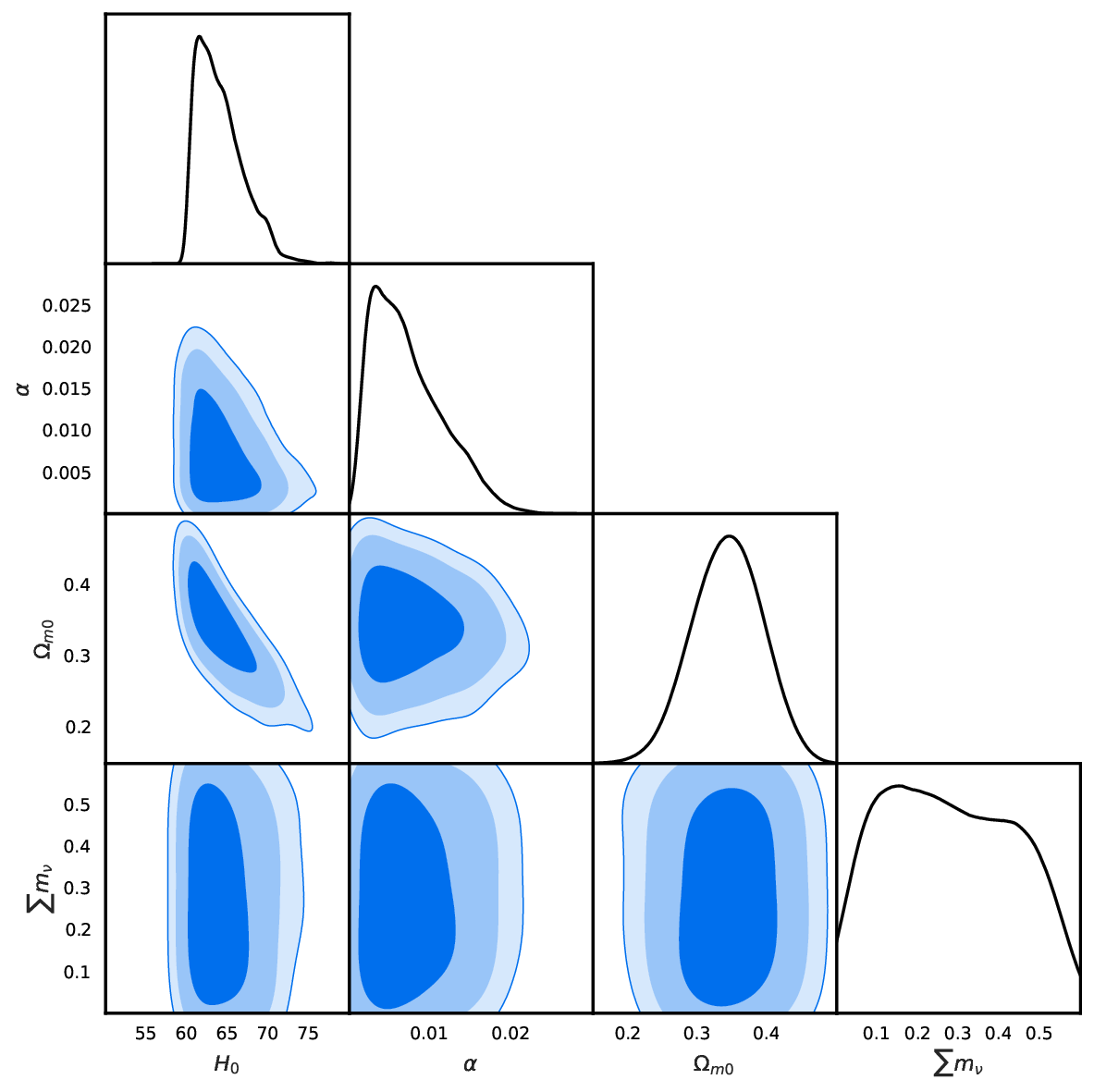}
\end{minipage}
\begin{minipage}{.49\textwidth}
\includegraphics[width=\linewidth]{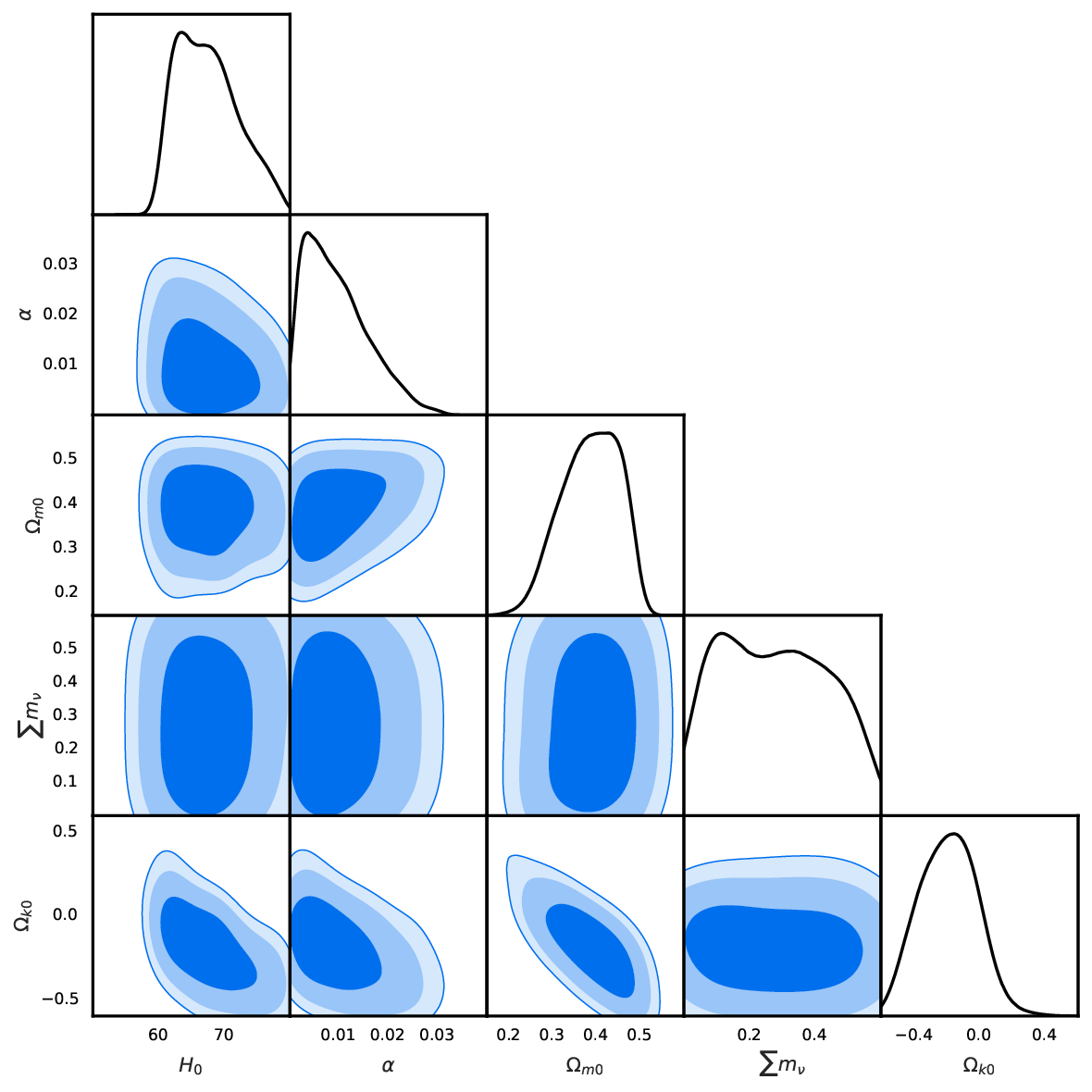}
\end{minipage}
\caption{One-dimensional likelihoods and $1\sigma$, $2\sigma$, and $3\sigma$ confidence level contours constraining the parameters of the MaVaN model from the $H(z)$ measurements: for flat space (left panel) and non-flat space (right panel).}
\label{fig:f6}
\end{figure*}

We carried out observational constraints on the parameters of the flat and non-flat MaVaN models, as well as on those of the standard $\Lambda$CDM model, using 32 $H(z)$ measurements and a MCMC analysis. The one-dimensional likelihoods and the $1\sigma$, $2\sigma$, and $3\sigma$ confidence level contours obtained from the $H(z)$ data for the flat MaVaN model are shown in Fig.~\ref{fig:f6} (left panel), while those for the non-flat MaVaN model are presented in Fig.~\ref{fig:f6} (right panel).

\begin{figure*}[!ht]
\centering
\begin{minipage}{.49\textwidth}
\includegraphics[width=\linewidth]{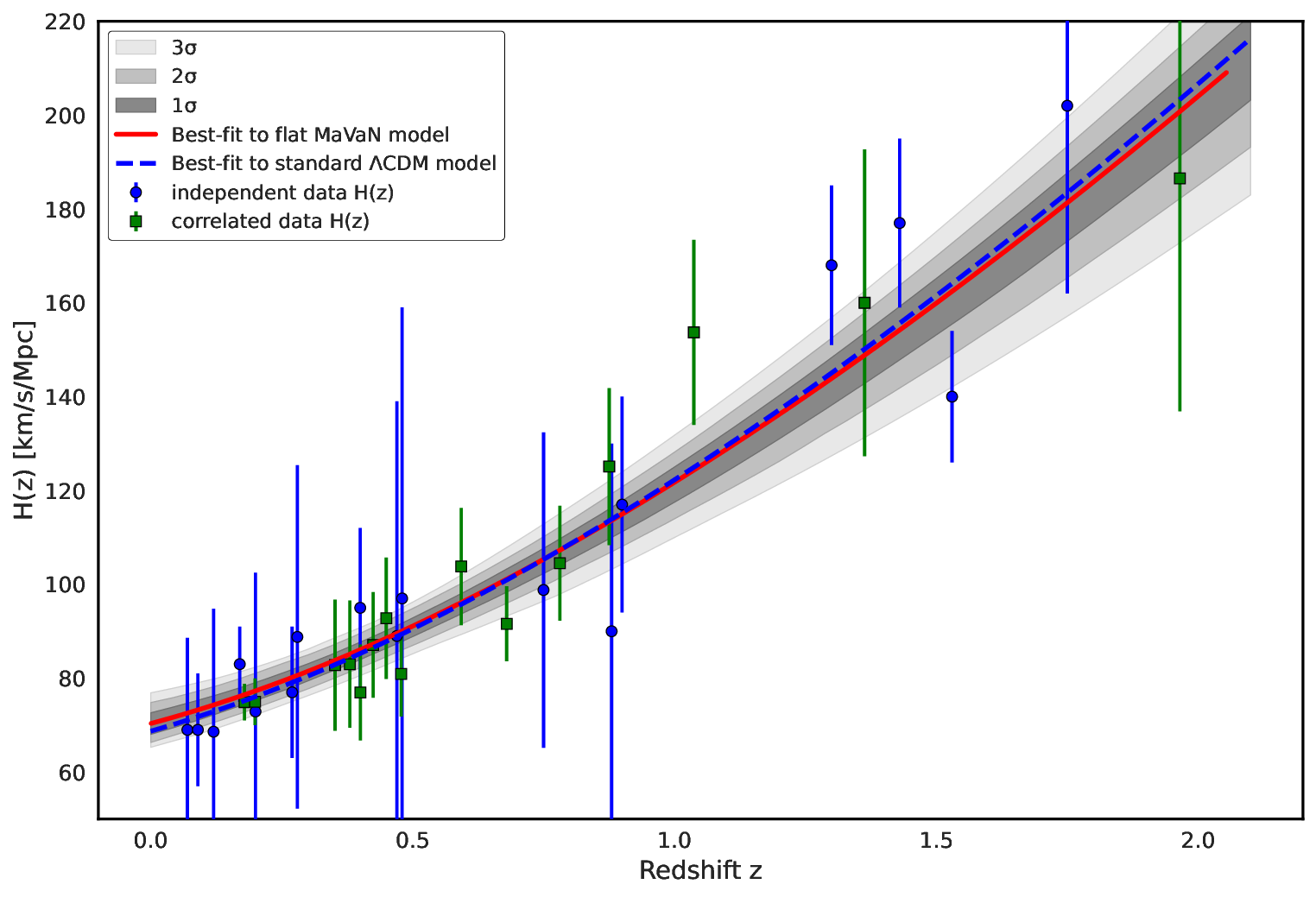}
\end{minipage}
\begin{minipage}{.49\textwidth}
\includegraphics[width=\linewidth]{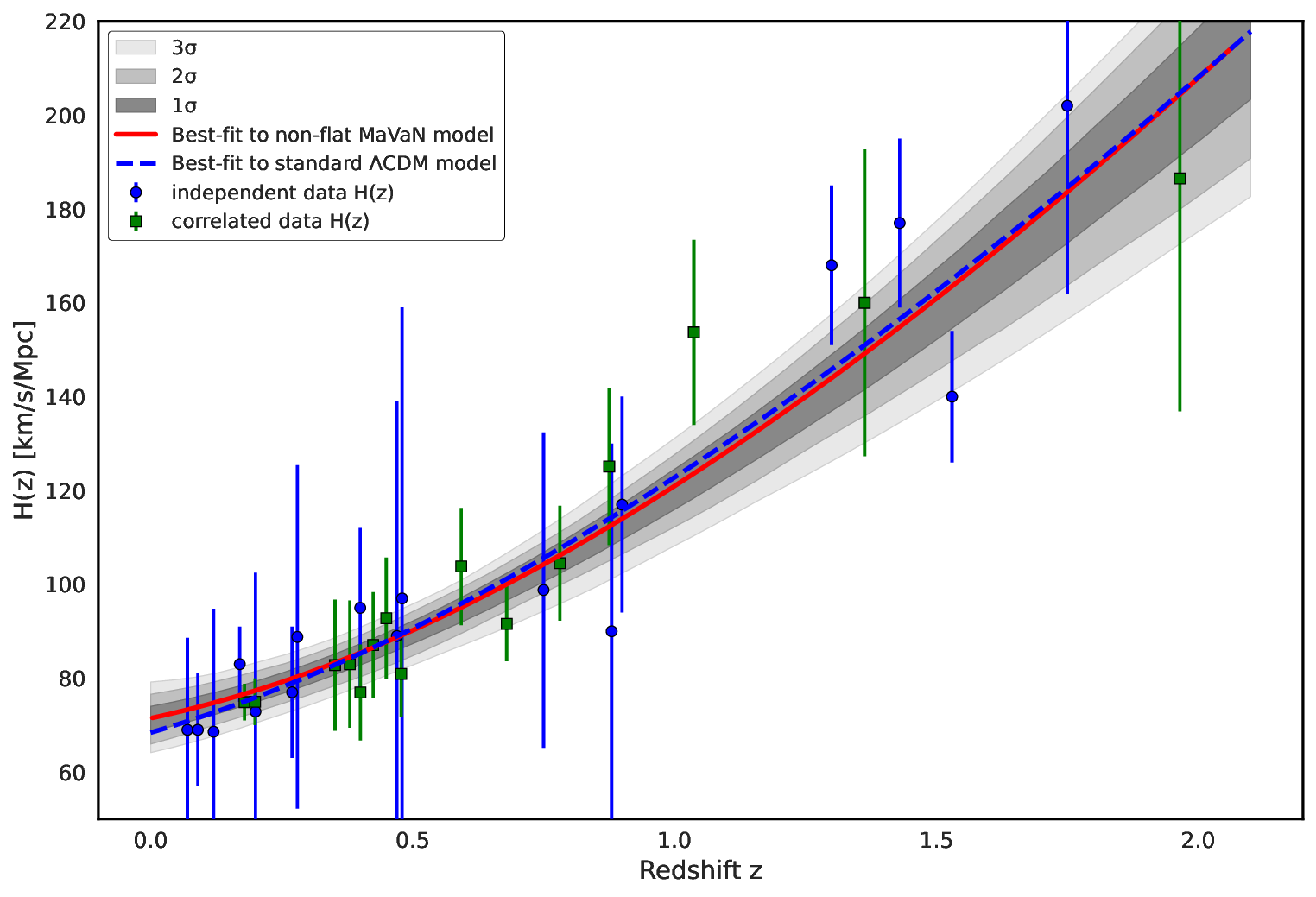}
\end{minipage}
\caption{Best-fit Hubble parameter values $H(z)$, with $1\sigma$, $2\sigma$, and $3\sigma$ confidence regions, and $1\sigma$ error bars for independent and correlated $H(z)$ data in the flat and non-flat MaVaN models, as well as in the standard $\Lambda$CDM model.}
\label{fig:f5}
\end{figure*}

We found that the $H(z)$ measurements have large uncertainties and almost all independent and correlated $H(z)$ data points, including their $1\sigma$ error bars, lie within the $3\sigma$ confidence region of the best-fit $H(z)$ predictions for both the flat and non-flat MaVaN models, as well as for the standard $\Lambda$CDM model (see Fig.~\ref{fig:f5}). 
The flat and non-flat MaVaN models introduce a mild redshift-dependent modification of the expansion history, resulting in small deviations from the $\Lambda$CDM model at low redshifts. These deviations change sign with redshift, which may be associated with the dynamical neutrino–scalar field coupling inherent to the MaVaN framework. However, the magnitude of these effects remains well below the $1\sigma$  level, indicating that there is no statistically significant deviation from the standard $\Lambda$CDM cosmological model.

The results of the constraints from observational $H(z)$ data on the free parameters of the flat and non-flat MaVaN models and the standard $\Lambda$CDM model are presented in Tables~(\ref{table:2}-\ref{table:3}). 
The non-flat MaVaN model predicts a slightly higher Hubble constant $H_0$ and a negative curvature parameter, indicating a slightly closed universe, while the flat MaVaN model gives a lower $H_0$. In the $\Lambda$CDM case, the geometry is fixed to be flat, and the resulting $H_0$ value is higher than in the flat MaVaN model and comparable to that obtained in the non-flat MaVaN model.
The parameters $\sum m_\nu$ and $\alpha$ are weakly constrained in both MaVaN models, and the $H(z)$ data do not favor significant deviations from zero. Furthermore, the $H(z)$ data are insufficient to place strong constraints  on the sum of neutrino mass $\sum m_\nu$ and the model parameter $\alpha$ within the MaVaN framework.

Based on $H(z)$ measurements, we investigated whether MaVaN cosmological models can mitigate the $H_0$ tension between early- and late-universe  measurements. The non-flat MaVaN model yields a value of $H_0$ consistent with the Planck CMB result, reducing the tension to a negligible level ($\lesssim 0.1\sigma$ confidence level) and providing a modest improvement relative to the $\Lambda$CDM model (from  $\sim 2\sigma$ to  $\sim 1.1\sigma$ confidence level). Due to the comparatively large uncertainties in the inferred $H_0$, the discrepancy with the SH0ES measurement also weakens to below the $1\sigma$ confidence level. 
In contrast, the flat MaVaN model predicts a substantially lower value of $H_0$, leading to a mild tension with Planck  CMB result at the $\sim 1.4\sigma$ confidence level and a stronger discrepancy with the SH0ES measurement at the $\sim 3\sigma$ confidence level, and is therefore disfavored by the data.
Overall, although the non-flat MaVaN scenario improves agreement with early-universe constraints, the apparent alleviation of the $H_0$ tension is primarily driven by large statistical uncertainties. Consequently, none of the models considered provides a robust resolution of the discrepancy with local measurements based solely on $H(z)$ data, and any apparent agreement with the SH0ES measurement should not be regarded as statistically conclusive. 

Using the $H(z)$ data, we compared the flat and non-flat MaVaN models with the standard $\Lambda$CDM model through the information criteria $AICc$, $BIC$, the differences $\Delta\chi^2_{\min}$, $\Delta AICc$, $\Delta BIC$, and the model probability weights $w_i$ and the reduced chi-square $\chi^2_{\rm min}/\mathrm{dof}$ (see Table~\ref{table:4}). The differences in $\chi^2_{\min}$ are small ($|\Delta\chi^2_{\min}|<1$) and all models provide comparably good fits, with $\chi^2_{\rm min}/\mathrm{dof} \ll 1$ reflecting the large uncertainties in the $H(z)$ data. The information criteria and probability weights strongly favor the standard $\Lambda$CDM model, which exhibits the lowest $AICc$ and $BIC$ values and a dominant weight of $w = 0.93$, whereas the flat and non-flat MaVaN models receive negligible weights ($w = 0.06$ and $0.02$), indicating that the additional parameters in the MaVaN scenarios are not justified by the current data.

These results indicate that, given the current $H(z)$ data, neither the flat nor the non-flat MaVaN model provides a statistically significant improvement over the standard $\Lambda$CDM model. The additional parameters in the MaVaN scenarios do not lead to a meaningful enhancement of the fit, highlighting the need for more precise observational data to robustly test such extended MaVaN cosmological models.

\section*{Acknowledgments}
This work was supported by Shota Rustaveli National Science Foundation of Georgia (SRNSFG) grant YS-22-998. The author thanks Bharat Ratra, Lado Samushia, Giorgi Khomeriki, Zack Brown, and Nick Magnelli for their hospitality in the Department of Physics, Kansas State University. The author thanks the anonymous referee for a careful reading of the manuscript and for valuable comments and suggestions that significantly improved the clarity and quality of this work.


\end{document}